\documentclass[12pt]{article}

\usepackage{graphics,amssymb,epsfig,float}
\usepackage[usenames,dvips]{color}
\usepackage{graphicx}
\usepackage{epsfig}
\usepackage{rotating}
\usepackage{dcolumn}
\usepackage{bm}
\usepackage{cite}
\usepackage{amsmath}

\def\lsim{\;\raise0.3ex\hbox{$<$\kern-0.75em\raise-1.1ex\hbox{$\sim$}}\;}
\def\gsim{\;\raise0.3ex\hbox{$>$\kern-0.75em\raise-1.1ex\hbox{$\sim$}}\;}
\def\beq{\begin{equation}}   \def\eeq{\end{equation}}
\def\ba{\begin{array}}       \def\ea{\end{array}}
\def\bea{\begin{eqnarray}}   \def\eea{\end{eqnarray}}
\def\nn{\nonumber}

\begin{document}

\begin{titlepage}
\begin{flushright}
LPT Orsay 11-125 \\ 
\end{flushright}

\begin{center}
\vspace{1cm}
{\Large\bf A Higgs boson near 125 GeV with enhanced di-photon signal in
the NMSSM} \\
\vspace{2cm}

{\bf{Ulrich Ellwanger}}
\vspace{1cm}\\
\it  Laboratoire de Physique Th\'eorique, UMR 8627, CNRS and
Universit\'e de Paris--Sud,\\
\it B\^at. 210, 91405 Orsay, France \\

\end{center}

\vspace{1cm}

\begin{abstract}
A natural region in the parameter space of the NMSSM can accomodate a
CP-even Higgs boson with a mass of about 125~GeV and, simultaneously, an
enhanced cross section times branching ratio in the di-photon
channel. This happens in the case of strong singlet-doublet mixing, when
the partial width of a 125~GeV Higgs boson into $b\,\bar{b}$ is strongly
reduced. In this case, a second lighter CP-even Higgs boson is
potentially also observable at the LHC.
\end{abstract}

\end{titlepage}

\section{Introduction}

Based on the analysis of 5 fb$^{-1}$ of data at the LHC, the ATLAS
\cite{:2012si} and CMS \cite{Chatrchyan:2012tx} collaborations have
presented evidence for a Higgs boson with a mass in the 125~GeV range.
The relevant search channels are $H \to \gamma\,\gamma$, $H \to Z\,Z^*
\to 4l$, $H \to W\,W^* \to 2l\,2\nu$ and to some extend (at CMS) $H \to
\tau\,\tau$. Interestingly, the best fit to the signal strength
$\sigma^{\gamma\gamma} = \sigma_{prod} \times BR(H \to \gamma\,\gamma)$
in the $\gamma\,\gamma$ search channel is by about one standard
deviation larger than expected in the Standard Model (SM) for both
collaborations: $\sigma^{\gamma\gamma}/\sigma_{SM}^{\gamma\gamma} \sim
2$ (ATLAS), and $\sigma^{\gamma\gamma}/\sigma_{SM}^{\gamma\gamma} \sim
1.7$ (CMS). Of course, the present evidence for a Higgs boson is not
(yet?) sufficiently significant in order to consider its existence as
assured, even less is the excess in the $H \to \gamma\,\gamma$ channel a
proof for a non-SM-like Higgs boson.

A relatively light Higgs boson (with a mass not too far above the LEP
bound of \linebreak $\sim114$~GeV) is a genuine prediction of
supersymmetric extensions of the SM which remain consistent up to a
Grand Unification (GUT) scale of about $10^{16}$~GeV, in particular in
the Minimal Supersymmetric extension (MSSM) with a minimal Higgs sector
consisting of two SU(2) doublets $H_u$ and $H_d$. In fact, in the MSSM
the solution of the fine tuning problem offered by supersymmetry works
the better, the lighter is the mostly SM-like Higgs boson.

Still, the parameter space of the MSSM allows to describe a Higgs boson
with a mass in the 125~GeV range if certain combinations of the stop
masses, stop mixings, $\tan\beta$ and the parameter $M_A$ (essentially
the heavy Higgs masses) are large enough  \cite{Gogoladze:2011aa,
1112.3017, 1112.3026, Arbey:2011ab,Arbey:2011aa,1112.3336,
Akula:2011aa,Kadastik:2011aa, Cao:2011sn,Arvanitaki:2011ck}. This
implies a fine tuning within the MSSM parameter space of the order of
1\% \cite{1112.2703}, or extra matter \cite{1112.3142}. An enhancement
of $\sigma^{\gamma\gamma}/\sigma_{SM}^{\gamma\gamma}$ may be possible in
the presence of light staus \cite{1112.3336}.

The Next-to-Minimal Supersymmetric Standard Model (NMSSM)
\cite{Ellis:1988er,Drees:1988fc,Ellwanger:2004xm,Maniatis:2009re,
Ellwanger:2009dp} is the simplest supersymmetric (Susy) extension of the
SM with a scale invariant superpotential, i.\,e. where the only
dimensionful parameters are the soft Susy breaking terms. No
supersymmetric Higgs mass term $\mu$ as in the MSSM is required, since
it is generated dynamically by the vacuum expectation value (vev) of a
gauge singlet superfield $S$ and a coupling $\lambda S H_u H_d$ in the
superpotential. Together with the neutral components of the two SU(2)
doublet Higgs fields $H_u$ and $H_d$ of the MSSM, one finds three
neutral CP-even Higgs states in this model. These three states mix in
the form of a $3 \times 3$ mass matrix and, accordingly, the physical
eigenstates are superpositions of the neutral CP-even components of
$H_u$, $H_d$ and $S$. In general, the couplings of the physical states
to gauge bosons, quarks and leptons can differ considerably from the
corresponding couplings of a SM Higgs boson. The possible alleviation in
the NMSSM of the ``little fine tuning problem'' in the Higgs sector of
the MSSM has been studied in \cite{Ellwanger:2011mu} in the light of 2
fb$^{-1}$ of data at the LHC, and in the light the recent evidence for a
Higgs mass of about 126~GeV in \cite{1112.2703} (although mostly for
large values of $\lambda$, implying new strong interactions below the
GUT scale).

In most of the parameter space of the NMSSM, the physical Higgs spectrum
contains a heavy CP-even state, a heavy CP-odd state and a charged Higgs
boson which are nearly degenerate as in the MSSM with a common mass
$\sim M_A$. However, the lighter doublet-like CP-even state
(corresponding to the SM-like Higgs boson $H_{SM}$) can mix
strongly with the real part of $S$ and form eigenstates with reduced
couplings to gauge bosons, quarks and leptons \cite{Ellis:1988er,
Drees:1988fc, Ellwanger:1993xa,Kamoshita:1994iv, Ellwanger:1995ru,
Franke:1995xn,King:1995ys, Ellwanger:1999ji, Ellwanger:2005uu,
Moretti:2006sv,Dermisek:2007ah, Ellwanger:2009dp,Ellwanger:2011sk}.
In this case, possibly both eigenstates are visible at the LHC (see
\cite{Ellwanger:2011sk}, where a second visible state with reduced
couplings in the $140 - 150$~GeV range has been studied, and refs.
therein).

It is well known that, for small values of $\tan\beta$, the coupling
$\lambda S H_u H_d$ in the superpotential leads to a positive
contribution to the mass squared of the SM-like Higgs boson $H_{SM}$
relative to the MSSM \cite{Ellis:1988er, Drees:1988fc,Ellwanger:2009dp}.
However, $H_{SM}-S$ mixing has an additional impact on the physical
spectrum: if the diagonal mass term $m_{SS}^2$ is larger than the one of
$H_{SM}$, the mixing reduces the mass of $H_{SM}$; if the diagonal mass
term $m_{SS}^2$ is smaller than the one of $H_{SM}$, the mixing leads to
an additional increase of the mass of $H_{SM}$. In this latter case, the
mass of the lighter eigenstate $H_1$ can be well below 114~GeV and
compatible with constraints from LEP \cite{Schael:2006cr}, if its
reduced signal strength $\xi_1^2 \equiv \bar{g_1}^2 \times
\overline{BR}(H_1 \to b \bar{b})$ is small enough. (Here $\bar{g_1}$ is
the reduced coupling of $H_1$ to the $Z$ boson normalized with respect
to the SM, and $\overline{BR}(H_1 \to b \bar{b})$ is the branching ratio
into $b \bar{b}$ normalized with respect to the SM.)

In addition, $H_{SM}-S$ mixing can lead to an increase of the branching
ratio $BR(H_i \to \gamma\,\gamma)$ of one of the eigenstates $H_i$ with
respect to the SM: if the coupling to $b\,\bar{b}$ and hence the partial
decay width into $b\,\bar{b}$ (which is close to the total width
$\Gamma_{Tot}$) is strongly reduced with respect to the SM, $BR(H_i \to
\gamma\,\gamma) = \Gamma(H_i \to \gamma\,\gamma)/ \Gamma_{Tot}$ is
correspondingly enhanced. This phenomenon has been discussed in the
context of the lighter eigenstate $H_1$ in \cite{Ellwanger:2010nf}, but
is equally possible for the heavier eigenstate as will be discussed
below. In view of the latest LHC results, the possible enhancement of
$BR(H_i \to \gamma\,\gamma)$ in the NMSSM was also discussed in
\cite{1112.2703}, and a Higgs mass near 125~GeV in the constrained NMSSM
-- but without enhancement of $BR(H_i \to \gamma\,\gamma)$ -- in
\cite{Gunion:2012zd}.

In the next Section we will study a region of the parameter space of the
NMSSM with a scale invariant superpotential, which leads naturally to an
eigenstate $H_2$ after $H_{SM}-S$ mixing with a mass in the $124 -
127$~GeV range. Its $BR(H_2 \to \gamma\,\gamma)$ is always enhanced with
respect to the SM. The lighter eigenstate $H_1$ has a mass in the $70 -
120$~GeV range, compatible with LEP constraints, and is potentially also
observable at the LHC. In Section~3 we conclude and summarize the
possibilities allowing to distinguish this scenario from the SM
and/or the MSSM.

\section{Implications of $H_{SM}-S$ mixing in the NMSSM in the light of
recent and future LHC results}

The NMSSM differs from the MSSM due to the presence of the gauge singlet
superfield $S$. In the simplest $Z_3$ invariant realisation of the
NMSSM, the Higgs mass term $\mu H_u H_d$ in the superpotential
$W_{MSSM}$ of the MSSM is replaced by the coupling $\lambda$ of $S$ to
$H_u$ and $H_d$ and a self-coupling $\kappa S^3$.  Hence, in this
simplest version the superpotential $W_{NMSSM}$ is scale invariant, and
given by:
\beq\label{eq:1}
W_{NMSSM} = \lambda \hat S \hat H_u\cdot \hat H_d + \frac{\kappa}{3} 
\hat S^3 + \dots\; ,
\eeq
where hatted letters denote superfields, and the dots denote the
MSSM-like Yukawa couplings of $\hat H_u$ and $\hat H_d$ to the quark and
lepton superfields. Once the real scalar component of $\hat S$ develops
a vev $s$, the first term in $W_{NMSSM}$ generates an effective
$\mu$-term
\beq\label{eq:2}
\mu_\mathrm{eff}=\lambda\, s\; .
\eeq
A constraint $|\mu_\mathrm{eff}| \gsim 100$~GeV follows from
the non-observation of higgsino-like charginos at LEP.

The soft Susy breaking terms consist of mass terms for the Higgs bosons
$H_u$, $H_d$ and $S$, and trilinear interactions (omitting squarks and
sleptons)
\beq\label{eq:3}
 -{\cal L}_\mathrm{Soft} =
m_{H_u}^2 | H_u |^2 + m_{H_d}^2 | H_d |^2 + 
m_{S}^2 | S |^2 +\Bigl( \lambda A_\lambda\, H_u
\cdot H_d \,S +  \frac{1}{3} \kappa  A_\kappa\,  S^3 \Bigl)+
\mathrm{h.c.}\; .
\eeq
Expressions for the mass matrices of the physical CP-even and CP-odd
Higgs states -- after $H_u$, $H_d$ and $S$ have assumed vevs $v_u$,
$v_d$ and $s$ and including the dominant radiative corrections -- can be
found in \cite{Ellwanger:2009dp} in will not be repeated here.
As compared to two independent parameters in the Higgs sector of the
MSSM at tree level (often chosen as $\tan \beta$ and $M_A$), the Higgs
sector of the NMSSM is described by the six parameters
\beq \label{eq:4}
\lambda\ , \ \kappa\ , \ A_{\lambda} \ , \ A_{\kappa}, \ \tan \beta\ =
v_u/v_d\ ,\ \mu_\mathrm{eff}\; .
\eeq
Alternatively, the parameter $A_{\lambda}$ can be replaced by the
MSSM-like parameter
\beq\label{eq:5}
M_{A}^2 = \frac{2\mu_\mathrm{eff} B_\mathrm{eff}}{\sin 2\beta}\; ,
\eeq
where $B_\mathrm{eff}= A_\lambda + \kappa s$.

Subsequently we are interested in regions of the parameter space where
the soft Susy breaking terms are not very large (in order to avoid large
fine tuning), but they have to comply with the present non-observation
of sparticles at the LHC. In the gaugino, squark and slepton sectors we
make the following choice, motivated to a certain extend by the
renormalization group running from the GUT scale down to the weak scale
(although the precise values are not very important): bino, wino and
gluino masses $M_1$=175~GeV, $M_2$=350~GeV and $M_3$=1000~GeV
respectively, squark masses of 1200~GeV (but 800~GeV for the third
generation), slepton masses of 300~GeV, $A_t = A_b = -1000$~GeV.

In the Higgs sector we have to keep in mind that the soft Susy breaking
masses $m_{H_u}^2$, $m_{H_d}^2$ and $m_{S}^2$ are determined implicitely
(through the minimization equations of the scalar potential) in terms of
$M_Z$, $\tan\beta$ and $\mu_\mathrm{eff}$. Large values of $m_{H_u}^2$,
$m_{H_d}^2$ and $m_{S}^2$ are avoided if $\mu_\mathrm{eff}$, $M_A$ and
$\tan\beta$ are relatively small. (Large values of $\tan\beta$ require
small tuned values for $B_\mathrm{eff}$ in the NMSSM, unless
$|m_{H_u}^2|$ and/or $|m_{H_d}^2|$ are large.) Hence we choose
$\mu_\mathrm{eff}=140$~GeV, $M_A=300$~GeV and $1.7 < \tan\beta < 2$
leading to $A_\lambda \sim 140 - 200$~GeV. Then, the interesting regions
of the remaining parameters $\lambda$, $\kappa$ and $A_\kappa$ are
determined by the conditions that i) one of the physical eigenstates in
the CP-even Higgs sector (actually always $H_2$) has a mass in the  $124
- 127$~GeV range, and ii) the lighter eigenstate $H_1$ is not in
conflict with LEP constraints. The density of viable points is
particularly large for $0.5 < \lambda < 0.6$, $0.3 < \kappa < 0.4$ and
$-250\ \mathrm{GeV}< A_\kappa < -200$~GeV. Of course, viable points
outside this range exist as well, but these will not invalidade our
subsequent conclusions.

A corresponding scan in parameter space is performed with the help of
the code NMSSMTools  \cite{Ellwanger:2004xm,Ellwanger:2005dv}; we
employed the version 3.0.2 which is includes radiative corrections to
the Higgs sector from \cite{Degrassi:2009yq}. Only points respecting
constraints on the Higgs sector from LEP and from B~physics are
retained. We find that about 50\% of all points in this region of
parameter space respect these phenomenological constraints, and $\sim
5-6\%$ ($\sim 550$ out of 10000) lead to a Higgs boson $H_2$ with a mass
in the $124 - 127$~GeV range. (Of course, measurements always reduce the
allowed regions in parameter space.)

The couplings of the Higgs states depend on their decompositions into
the CP-even weak eigenstates $H_d$, $H_u$ and $S$, which are given by
\bea\label{eq:6}
H_1 &=& S_{1,d}\ H_d + S_{1,u}\ H_u +S_{1,s}\ S\; ,\nn \\
H_2 &=& S_{2,d}\ H_d + S_{2,u}\ H_u +S_{2,s}\ S\; .
\eea
Then the reduced tree level couplings (relative to a SM-like Higgs
boson) of $H_i$ to $b$ quarks, $\tau$ leptons, $t$ quarks and
electroweak gauge bosons $V$ are
\bea
\frac{g_{H_i bb}}{g_{H_{SM} bb}} = \frac{g_{H_i \tau\tau}}{g_{H_{SM}
\tau\tau}} &=& \frac{S_{i,d}}{\cos\beta}\;,\qquad
\frac{g_{H_i tt}}{g_{H_{SM} tt}} = \frac{S_{i,u}}{\sin\beta}\;,\nn \\
\bar{g}_i \equiv \frac{g_{H_i VV}}{g_{H_{SM} VV}} &=& \cos\beta\,
S_{i,d} + \sin\beta\, S_{i,u}\; .
\label{eq:7}
\eea

For the low values of $\tan\beta$ considered here, the couplings of
Higgs bosons to gluons (relevant for their production) and to photons
are induced by loop diagrams dominated by top-quark loops. As stated
above, the branching ratios into two photons can be enhanced, if the
coupling to $b$-quarks is reduced, which is the case if $S_{i,d}$ is
small. 

Subsequently we are interested in the signal strength
$\sigma_2^{\gamma\gamma} = \sigma_{prod} \times BR(H_2 \to
\gamma\,\gamma)$ relative to the SM, $R_2^{\gamma\gamma} =
\sigma_2^{\gamma\gamma}/\sigma_{SM}^{\gamma\gamma}$.
$R_2^{\gamma\gamma}$ is the product of two factors: i) the reduced
coupling of $H_2$ to gluons, which is essentially given by ${g_{H_2
tt}}/{g_{H_{SM} tt}}$ (but contributions from non-SM particles in the
loop are taken into account), and ii) the $\overline{BR}(H_2 \to
\gamma\,\gamma)$, the branching ratio of $H_2$ into $\gamma\,\gamma$
normalized with respect to the corresponding branching ratio of a
SM-like Higgs boson of the same mass.\footnote{The branching ratios of
SM-like Higgs bosons are computed in a subroutine hadecay.f within
NMSSMTools, such that radiative corrections are included at the same
level of accuracy. In fact, QCD corrections cancel in the ratio of
branching ratios NMSSM/SM. The routines for the branching ratios of
NMSSM- and SM-like Higgs bosons are based on modified versions of HDECAY
\cite{Djouadi:1997yw}.}
$\overline{BR}(H_2 \to
\gamma\,\gamma)$ can be considerably larger than~1. In Fig.~1 we show
$R_2^{\gamma\gamma}$ as function of $S_{2,d}^2$ for $\sim 550$ points in
the region of the parameter space of the NMSSM described above, in which
$M_{H_2}$ is in the $124 - 127$~GeV range. We see that
$R_2^{\gamma\gamma}$ is \emph{always} larger than 1.1, with an expected
dependence on $S_{2,d}^2$.

If one modifies somewhat the soft Susy breaking squark and slepton
masses (and trilinear couplings $A$) at the weak scale, the parameters
can be mapped to a semi-constrained version of the NMSSM together with
non-universal soft Higgs masses at the GUT scale as studied in
\cite{Gunion:2012zd} (where additional regions in the parameter space of
the NMSSM with $M_{H_1}$ or $M_{H_2} \sim 124 \dotsc 127$~GeV have been
found). One obtains $M_{1/2}\sim 500$~GeV, $m_0 \sim 500\dots 700$~GeV,
$A_0 \sim -900 \dotsc -950$~GeV, for the soft terms involving the
singlet $m_S \sim 1$~TeV, $A_\lambda \sim -400$~GeV, $A_\kappa \sim
-300$~GeV, and for the soft Higgs masses $m_{H_u} \sim 1.5$~TeV,
$m_{H_d} \sim m_0$. Due to the low value of $\tan\beta$ (leading to a
relatively large value of the top Yukawa coupling $h_t$) and the large
values of $\lambda$, $\kappa$ at the weak scale, all 3~Yukawa couplings
are of ${\cal O}(1)$ at the GUT scale: $h_t \sim 1.2 \dotsc 1.3$,
$\lambda \sim 1.3 \dotsc 1.7$, $\kappa \sim 0.7 \dotsc
1.0$.\footnote{For such large Yukawa couplings, the solution of
renormalization group equations for the running parameters with boundary
conditions both at the weak \emph{and} the GUT scale is a delicate issue
leading to convergence problems. In fact, the public version of the code
NMSPEC inside NMSSMTools \cite{Ellwanger:2006rn} has to be modified for
a study of this region.} The fact that all 3~Yukawa couplings are close
to (but just below) a Landau singularity at the GUT scale is intriguing.

\begin{figure}[ht!]
\begin{center}
\includegraphics*[scale=0.6,angle=-90]{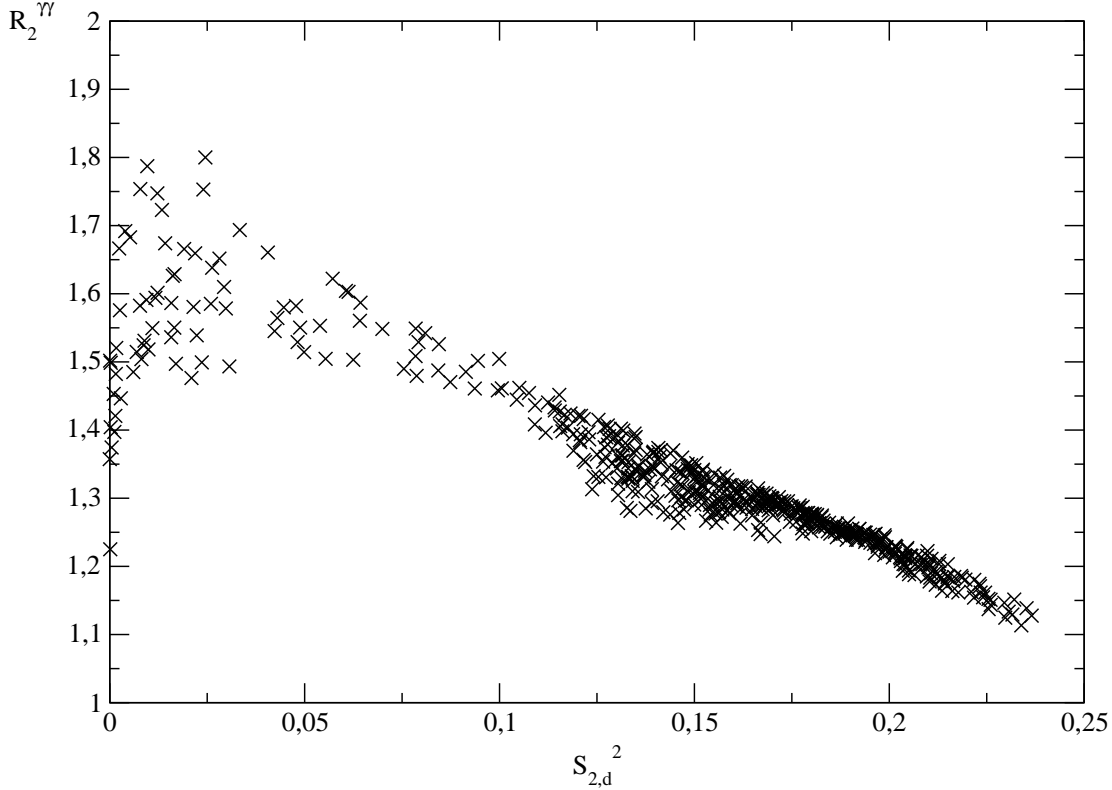}
\caption{The relative signal rate $R_2^{\gamma\gamma} =
\sigma_2^{\gamma\gamma}/\sigma_{SM}^{\gamma\gamma}$ as function of
$S_{2,d}^2$, for $H_2$ with a mass in the $124 - 127$~GeV range for
about 550 points in the parameter space of the NMSSM described in the
text}
\label{fig:1}
\end{center}
\end{figure}

Next we turn to the lighter Higgs boson $H_1$ in this scenario. Its mass
is in the $70 - 120$~GeV range. The most relevant search channels in
this mass range are again the $\gamma\,\gamma$ mode, but also $H_1 \to
\tau\,\tau$ (with $H_1$ produced by vector boson fusion, VBF) and, to
some extent, $H_1 \to b\,\bar{b}$ with $H_1$ produced in association
with $W$ or $Z$ bosons. The reduced signal strength in the
$\gamma\,\gamma$ mode, $R_1^{\gamma\gamma} =
\sigma_1^{\gamma\gamma}/\sigma_{SM}^{\gamma\gamma}$, can be obtained as
above. The reduced signal strength in the $\tau\,\tau$ mode and VBF,
$R_1^{\tau\,\tau} = \sigma_1^{\tau\tau}/\sigma_{SM}^{\tau\tau}$, is the
product of the reduced coupling $\bar{g}_1^2$ of $H_1$ to the
electroweak gauge bosons, and the $\overline{BR}(H_1 \to \tau\,\tau)$,
the branching ratio of $H_1$ into $\tau\,\tau$ normalized with respect
to the corresponding branching ratio of a SM-like Higgs boson of the
same mass. (The reduced signal strength in the $b\,\bar{b}$ mode is
practically the same as $R_1^{\tau\tau}$, since it is again
proportional to the coupling to electroweak gauge bosons, and the
branching ratio into $b\,\bar{b}$ remains proportional to the branching
ratio into $\tau\,\tau$.)

In Fig.~2 we show $R_1^{\gamma\gamma}$ and $R_1^{\tau\,\tau}$ as
function of $M_{H_1}$. We see that $R_1^{\gamma\gamma}$ is not enhanced,
but mostly strongly reduced due to the small coupling of $H_1$ to two
gluons, which is \emph{not} compensated by an enhanced branching ratio
into two photons in this case. Hence, except perhaps for $M_{H_1} \gsim
110$~GeV, the prospects for a discovery of $H_1$ in this channel are not
rosy. Likewise, $R_1^{\tau\tau}$ ($\simeq R_1^{b\bar{b}}$) is not
enhanced, but not as small as $R_1^{\gamma\gamma}$. Actually the upper
bound on $R_1^{\tau\tau}$ coincides with the upper LEP bound on $\xi_1^2
\equiv \bar{g_1}^2 \times \overline{BR}(H_1 \to b \bar{b})$ as function
of $M_H$ \cite{Schael:2006cr}, which is not astonishing given that
$\overline{BR}(H_1 \to b\, \bar{b}) \sim \overline{BR}(H_1 \to
\tau\,\tau)$. Hence, although a discovery of $H_1$ in the $\tau\,\tau$
channel (or $b\,\bar{b}$ mode) is not guaranteed, this is not excluded
in particular after future high luminosity runs of the LHC or if its
mass is in the $110 - 120$~GeV range.

\begin{figure}[t!]
\begin{center}
\includegraphics*[scale=0.6,angle=-90]{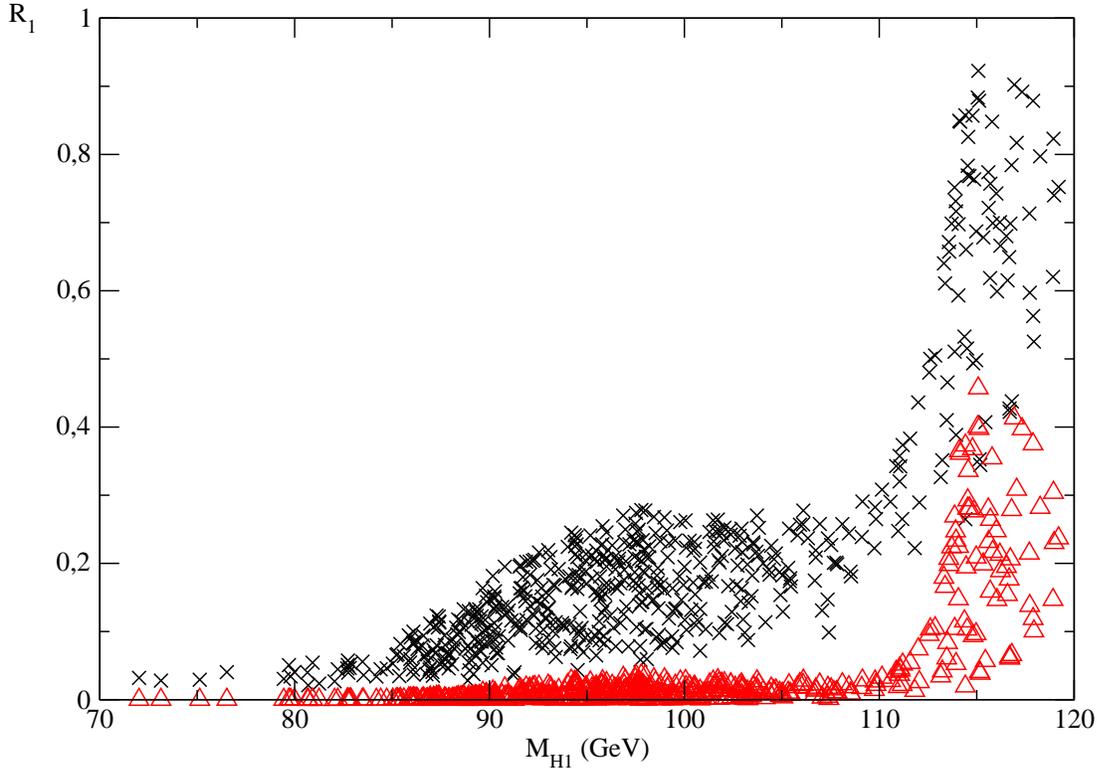}
\caption{The relative signal rate $R_1=R_1^{\gamma\gamma} =
\sigma_1^{\gamma\gamma}/\sigma_{SM}^{\gamma\gamma}$ (red triangles) and
$R_1=R_1^{\tau\tau} = \sigma_1^{\tau\tau}/\sigma_{SM}^{\tau\tau}$
(black crosses) as function of $M_{H_1}$, for about 550 points in the
parameter space of the NMSSM described in the text}
\label{fig:2}
\end{center}
\end{figure}

\section{Conclusions}

We have presented a natural region in the parameter space of the NMSSM,
where the NMSSM-specific coupling $\lambda$ and mixing effects push up
the mass of a CP-even Higgs boson into the $124 - 127$~GeV range without
the need for excessive radiative corrections from heavy sparticles. The
relative signal rate in the $\gamma\,\gamma$ channel is always enhanced
by a factor 1.1 -- 1.8 with respect to a SM-like Higgs boson of the same
mass. This Higgs boson complying with recent evidence from the ATLAS and
CMS collaborations is accompagnied by a lighter CP-even neutral Higgs
state.

Under the following circumstances it might be possible to distinguish
this scenario from the SM and/or the MSSM:
\begin{itemize}
\item the enhanced signal rate in the $\gamma\,\gamma$ channel is
confirmed, and incompatible with a SM-like Higgs boson;
\item sparticles are detected, and their masses turn out to be
incompatible with the necessarily large radiative corrections to the
Higgs mass in the MSSM;
\item the lighter CP-even state $H_1$ is discovered.
\end{itemize}

Of course, first of all the present evidence for a Higgs boson into the
$124 - 127$~GeV range should be confirmed by more data; then the same
data can give us possible hints for non-SM-like properties of the Higgs
sector along the lines discussed here.

\section*{Acknowledgements}

The author acknowledges support from the French ANR LFV-CPV-LHC.


\end{document}